\documentclass[aps,pra,amssymb,twocolumn,superscriptaddress,showpacs]{revtex4}

\usepackage{graphicx}
\usepackage{dcolumn}
\usepackage{bm}
\usepackage{color}
\usepackage{epstopdf}
\usepackage{dsfont}

\begin{document}
\voffset1cm

\newcommand{\beq}{\begin{equation}}
\newcommand{\eeq}{\end{equation}}
\newcommand{\barr}{\begin{eqnarray}}
\newcommand{\earr}{\end{eqnarray}}

\newcommand{\rev}[1]{{\color{red}#1}}
\newcommand{\REV}[1]{\textbf{\color{red}[[#1]]}}
\newcommand{\BLUE}[1]{\textbf{\color{blue}#1}}
\newcommand{\GREEN}[1]{\textbf{\color{green}#1}}

\newcommand{\andy}[1]{ }
\newcommand{\bmsub}[1]{\mbox{\boldmath\scriptsize $#1$}}

\def\R{\mathbb{R}}

\def\bra#1{\langle #1 |}
\def\ket#1{| #1 \rangle}
\def\sinc{\mathop{\text{sinc}}\nolimits}
\def\cV{\mathcal{V}}
\def\cH{\mathcal{H}}
\def\cT{\mathcal{T}}
\def\cM{\mathcal{M}}
\def\CW{\mathcal{W}}
\def\CR{\mathcal{R}}
\renewcommand{\Re}{\mathop{\text{Re}}\nolimits}
\newcommand{\tr}{\mathop{\text{Tr}}\nolimits}
\def\e{\mathrm{e}}
\def\ii{\mathrm{i}}
\def\d{\mathrm{d}}

\title{Generalized  tomographic maps and star-product formalism}

\author{M. Asorey}
\affiliation{Departamento de F\'\i sica Te\'orica, Facultad de
Ciencias, Universidad de Zaragoza, 50009 Zaragoza, Spain}
\author{P. Facchi}
\affiliation{Dipartimento di Fisica and MECENAS, Universit\`a di Bari, I-70126 Bari, Italy} 
\affiliation{INFN, Sezione di Bari, I-70126 Bari, Italy}
\author{V.I. Man'ko}
\affiliation{P.N. Lebedev Physical Institute, Leninskii Prospect
53, Moscow 119991, Russia}
\author{G. Marmo} \affiliation{Dipartimento di Fisica and MECENAS, 
Universit\`a di Napoli ``Federico II", I-80126  Napoli, Italy}
\affiliation{INFN, Sezione di Napoli, I-80126  Napoli, Italy}
\author{S. Pascazio}
\affiliation{Dipartimento di Fisica and MECENAS, Universit\`a di Bari, I-70126 Bari, Italy} 
\affiliation{INFN, Sezione di Bari, I-70126 Bari, Italy}
\author{E.C.G. Sudarshan} \affiliation{Department of Physics,
University of Texas, Austin, Texas 78712, USA}

\date{\today}

\begin{abstract}
We elaborate on the notion of generalized tomograms, both in the classical and quantum domains. We construct a scheme of star-products of thick tomographic symbols and obtain in explicit form the kernels of classical and quantum generalized tomograms. Some of the new tomograms may have interesting applications in quantum optical tomography.
\end{abstract}


\maketitle

\section{Introduction}

The tomographic probability description of classical and quantum states \cite{NuovoCim,Ibort2} is based on the standard Radon transform 
\cite{Rad1917} of positive definite functions and Wigner functions \cite{Wig32, Ber-Ber, Vog-Ris}, respectively, the latter being actually experimentally used for the reconstruction of quantum states, through the determination of the associated density operators \cite{SBRF93,lwry, Mlynek96,konst} (see
also \cite{torino, ps,bellini,allevi09}) . 

The use of the Radon transform to associate a probability distribution (tomogram) with every Wigner function \cite{Ber-Ber,Vog-Ris}
was shown to be equivalent \cite{MarmoJPA} to the use of a specific scheme of quantizer-dequantizer procedure to build  a star-product
by means of a quantum version of the Radon transform. Specifically, the kernel of the associative star-product of tomographic symbols of quantum observables and the kernel of the associative star-product of symbols associated with classical observables were studied in \cite{NuovoCim,Ibort2,MarmoJPA}.

The aim of this paper is to further develop the star-product scheme by means of the so called {\it thick tomography} approach \cite{tomothick} for the description of quantum states and obtain in explicit form the star-product kernel for 
{\it thick tomographic} symbols. We also provide a general approach for the construction of tomographic schemes where
different nonlinear functions (not only linear ones) in position and momentum are being used, and take the opportunity to better qualify a statement made in Ref. \cite{quantomogram}. 

This article is organized as follows. In section 2, we review the general scheme for the construction of a star-product. In section 3, the thick tomographic approach is considered by following Ref.\ \cite{tomothick}, and the kernel for the star-product is constructed. 
In section 4, a general scheme to deal with nonlinear functions of position and momentum is formulated. Perspectives and conclusions are provided in section 5.

\section{General scheme of star-products}
The Weyl-Wigner formalism enables one to represent operators by means of functions on phase space and viceversa, in a one-to-one correspondence, provided that the functions satisfy appropriate conditions. By this one-to-one correspondence it is possible to build a non-local product on functions which corresponds to the associative operator product. As the Weyl map realizes a projectively unitary representation of the Abelian vector group (phase-space), the induced product on functions coincides with a twisted convolution product. This construction can be extended and generalized to any measure space $X$ on which it is possible to define, for any $x\in X$ a pair of operators $\hat{U}_1(x)$ and $\hat{D}_1(x)$ acting on some Hilbert space $\cal{H}$, called dequantizer and quantizer respectively, with the property 
\begin{equation}
\tr \hat{D}_1({x}) \hat{U}_1({x}') = \delta_{x'}(x),
\end{equation}
where $x\in X$.
To any operator $\hat{A}$ acting on $\cal{H}$ one can  associate a function called its symbol
\begin{equation}
A_1({x}) = \tr \hat{A}\hat{U}_1({x}),
\end{equation}
and the reconstruction formula, the inverse map,  reads
\begin{equation}
\hat{A} = \int A_1({x}) \hat{D}_1({x}) \d{x}.
\end{equation}
By using this one-to-one map we define  the star product of two symbols associated with $\hat{A}$ and $\hat{B}$ by
setting
\begin{eqnarray}
A_1 \star B_1 ({x}) &=& \tr \hat{A} \hat{B} \hat{U}_1({x}) 
\nonumber\\
&=& \int K_1({x}_1,{x}_2,{x}) A_1({x_1}) B_1({x_2}) \d{x}_1 \d{x}_2,
\quad
\end{eqnarray}
where the kernel of the star product is given by the relation~\cite{MarmoJPA}
\begin{equation}
K_1({x}_1,{x}_2,{x}) = \tr \left(\hat{D}_1({x}_1)\hat{D}_1({x}_2) \hat{U}_1({x}) \right).
\end{equation}
Let us assume that there exists another manifold (measure space) $Y$ with another pair of quantizer and dequantizer
operators $\hat{D}_2({y})$ and  $\hat{U}_2({y})$ with $y\in Y$, 
acting on the same Hilbert space $\cal{H}$
on which  $\hat{D}_1({x})$ and $\hat{U}_1({x})$ act. Then, with the same operator $\hat{A}$ it is possible to associate another symbol function defined on $Y$, say
\begin{equation}
A_2({y}) = \tr \hat{A}\hat{U}_2({y}),
\end{equation}
with inverse 
\begin{equation}
\label{vn}
\hat{A} = \int A_2({y}) \hat{D}_2({y}) \d{y}.
\end{equation}
By using formula (\ref{vn}), we find
\begin{equation}
\,\!\!\!\! \!\! A_1({x})\! 
=\!\!\! \int_Y \!\!K_{12}({x},{y}) A_2({y}) \d{y},
\label{eq:G7}
\end{equation}
\begin{equation}
\,\!\!\!\! \!\! A_2({y})\! 
=\!\!\! \int_X \!\!K_{21}({y},{x}) A_1({x}) \d{x},
\label{eq:G8}
\end{equation}
where we have defined the integral transform from symbols on $X$ to symbols on $Y$, and viceversa, by means of the formulae
\begin{equation}
K_{12}({x},{y}) = \tr \hat{U}_1({x})\hat{D}_2({y}),
\label{eq:G9}
\end{equation}
\begin{equation}
K_{21}({y},{x}) = \tr \hat{U}_2({y})\hat{D}_1({x}).
\label{eq:G10}
\end{equation}
The previous construction raises a natural question: given the integral relations (\ref{eq:G7})-(\ref{eq:G8}), is it possible
to find their quantum descriptions, that is, to find two pairs $\hat{U}_1(x)$ and $\hat{D}_1(x)$ and 
$\hat{U}_2(y)$ and $\hat{D}_2(y)$ such that (\ref{eq:G9}) and (\ref{eq:G10}) hold? This inverse problem, i.e., to find the quantizer-dequantizer pair corresponding to a given kernel was considered in \cite{Ibort2}. Now we will show how the standard symplectic tomography can be framed in this context. As a matter of fact, the classical  Radon transform and its inverse play 
the role of $K_{12}$ and $K_{21}$.

To show this explicitly, set  ${x}=(X,\mu,\nu)\in \mathbb{R}^3$ and ${y}=(q,p)\in\mathbb{R}^2$. Then for two functions $A_1({x})$ and $A_2({y})$ one has
\begin{equation}
A_1(X,\mu,\nu) = %\frac{1}{2\pi} 
\int A_2(q,p)\, \delta(X-\mu q-\nu p) \,\d q \d p,
\label{eq:G11}
\end{equation}
\begin{equation}
A_2(q,p)= \frac{1}{4\pi^2} \int A_1(X,\mu,\nu)\, \e^{\ii (X-\mu q- \nu p)} \,\d X \d\mu\d\nu.
\label{eq:G12}
\end{equation}
We now have to exhibit two quantizer-dequantizer pairs such that the kernels of~(\ref{eq:G11}) and (\ref{eq:G12}) have the form~(\ref{eq:G9}) and (\ref{eq:G10}). The solution is known and is provided by 
\begin{eqnarray}
\hat{U}_1(X,\mu,\nu) & = & \delta(X-\mu\hat{q}-\nu\hat{p}), \nonumber\\
\hat{D}_1(X,\mu,\nu) & = & \frac{1}{4\pi^2}\exp \ii (X-\mu\hat{q}-\nu\hat{p}),
\end{eqnarray}
and
\begin{eqnarray}
\hat{U}_2(q,p) & = & \int \e^{-\ii p u} \ket{q-\frac{u}{2}}\bra{q+\frac{u}{2}}\d u, \nonumber\\
\hat{D}_2(q,p) & = & \frac{1}{2\pi} \hat{U}_2(q,p).
\end{eqnarray}
The previous formulae show that the standard tomographic picture may be given a quantum version, indeed the {\it classical}
symbols are originated from {\it quantum} operators.

\section{Thick Tomography}

One can  use  for ``thick" quantizer and dequantizer for an arbitrary window function $\Xi(Y)\geq 0$.
\begin{equation}
\hat{U}(X,\mu,\nu) = \Xi(X-\mu \hat{q}-\nu\hat{p}),
\end{equation}
\begin{equation}
\hat{D}(X,\mu,\nu)=
\frac{\mathcal{N}_{\Xi}}{4\pi^2} \,
\e^{\ii (X -\mu\hat{q}-{\nu}\hat{p})},
\label{qsim}
\end{equation}
where
\begin{equation}
\mathcal{N}_{\Xi}= \left( \int {\Xi}(z)  \, {\rm e}^{\ii z}\, \d{z} \right)^{-1}.
\end{equation}

The kernel of the star product reads
\begin{eqnarray}
& & K_{\Xi}(X_1,\mu_1,\nu_1,X_2,\mu_2,\nu_2,X_3,\mu_3,\nu_3) 
\nonumber\\
&=&  \tr\left\{\hat{D}(X_1,\mu_1,\nu_1) \hat{D}(X_2,\mu_2,\nu_2)\hat{U}(X_3,\mu_3,\nu_3)\right\}
\nonumber\\
&= & \left(\frac{\mathcal{N}_{\Xi}}{2\pi}\right)^2 \e^{\ii X_1 + \ii X_2}
\nonumber\\
& & \tr\left\{\e^{-\ii \mu_1 \hat{q}-\ii \nu_1 \hat{p}}
\e^{-\ii \mu_2 \hat{q}-\ii \nu_2 \hat{p}}
\Xi(X_3-\mu_3 \hat{q}-\nu_3\hat{p})\right\}. \quad
\end{eqnarray}
Using the relation
\begin{equation}
\Xi(X) = \int \Xi(Y) \, \delta(X-Y) \, \d Y ,
\end{equation}
we get
\begin{eqnarray}
& & K_{\Xi} (X_1,\mu_1,\nu_1,X_2,\mu_2,\nu_2,X_3,\mu_3,\nu_3)
\nonumber\\
 & &=  \int K_{\delta} (X_1,\mu_1,\nu_1,X_2,\mu_2,\nu_2,X_3-Y,\mu_3,\nu_3)\, \Xi(Y) \, \d Y,
 \nonumber\\
\end{eqnarray}
where $K_{\delta}$ is the kernel of the star product related to the ideal symplectic tomogram (formally obtained when $\Xi=\delta$) \cite{MarmoJPA}.

Exactly the same relation takes place for classical commutative tomographic star product. It means that quantum and classical tomographic products are related by a twist factor, namely,
\begin{equation}
K_{\mathrm{qu}} = \e^{\frac{\ii}{2}(\nu_1\mu_2-\nu_2 \mu_1)} K_{\mathrm{cl}}.
\end{equation}

\section{Generalized tomographic star-product scheme in classical and quantum pictures}

Here we present a general scheme that relates the known classical tomographic schemes, given e.g.\ in~\cite{tomothick,tomogram} with their quantum versions. In fact, we apply the Weyl quantization map to formulate the integral relations among functions on phase space of a system with $n$ degrees of freedom. Let us consider two arbitrary functions: $f(\bm{q},\bm{p})$ on phase space $(\bm{q},\bm{p})\in\mathbb{R}^{2n}$, and  $\mathcal{W}_f(x)$ on an $m$-dimensional manifold $x\in X$. Assume that there are two integral relations between the functions:
\begin{eqnarray}
\mathcal{W}_f(x) & = & \int_{\mathbb{R}^{2n}} f(\bm{q},\bm{p}) \varphi(\bm{q},\bm{p},x) \d \bm{q} \d \bm{p},
\label{eq:rel1} \\
f(\bm{q},\bm{p})  & = & \int_X \mathcal{W}_f (x) \chi(\bm{q},\bm{p},x) \d x,
\label{eq:rel2}
\end{eqnarray}
with kernels $\varphi$ and $\chi$ that relate the phase-space with the manifold $X$.
In order to get a quantum version of these relations, we introduce the operators
\begin{eqnarray}
\hat{\varphi}(\hat{\bm{q}}, \hat{\bm{p}}, x) & = & \int_{\mathbb{R}^{2n}} \varphi(\bm{q},\bm{p},x) \hat{D}(\bm{q},\bm{p}) \d \bm{q} \d \bm{p},  \label{eq:rel3} \\
\hat{\chi}(\hat{\bm{q}}, \hat{\bm{p}}, x) & = & \int_{\mathbb{R}^{2n}} \chi(\bm{q},\bm{p},x) \hat{D}(\bm{q},\bm{p}) \d \bm{q} \d \bm{p}.
\label{eq:rel4}
\end{eqnarray}
The operators act on the Hilbert space of an $n$-mode system, e.g.\ on the space of states of an $n$-dimensional harmonic oscillator $L^2(\mathbb{R}^n)$. The  components of the operator vectors $\hat{\bm{q}}=(\hat{q}_1,\hat{q}_2,\dots,\hat{q}_n)$ and $\hat{\bm{p}}= (\hat{p}_1,\hat{p}_2,\dots,\hat{p}_n)$ are the standard position and momentum operators with commutators $[\hat{q}_k,\hat{p}_k]=\ii$ (with units $\hbar=1$). The explicit form of the operator $\hat{D}(\bm{q},\bm{p})$ is
\begin{equation}
\hat{D}(\bm{q},\bm{p}) = \frac{1}{\pi^n} \exp(2 \bm{\alpha} \cdot \hat{\bm{a}}^\dagger - 2 \bm{\alpha}^* \cdot \hat{\bm{a}}) \hat{I} ,
\end{equation}
where $\bm{\alpha} = (\alpha_1,\alpha_2, \dots, \alpha_n) \in \mathbb{C}^n$, with $\bm{\alpha} = (\bm{q} + \ii \bm{p})/\sqrt{2}$, while $\hat{\bm{a}}=(\hat{a}_1,\hat{a}_2,\dots,\hat{a}_n)$ and $\hat{\bm{a}}^\dagger=(\hat{a}^\dagger_1,\hat{a}^\dagger_2,\dots,\hat{a}^\dagger_n)$ are the vectors of annihilation and creation operators: $\hat{\bm{a}}= (\hat{\bm{q}} + \ii \hat{\bm{p}})/\sqrt{2}$ and $\hat{\bm{a}}^\dagger= (\hat{\bm{q}} - \ii \hat{\bm{p}})/\sqrt{2}$. The operator $\hat{I}$ is the parity operator with action $\hat{I} \psi(\bm{q}) = \psi(-\bm{q})$, for $\psi\in L^2(\mathbb{R}^n)$. The operator $\hat{D}(\bm{q},\bm{p})$ is the Weyl system displacement operator on the Hilbert space of an $n$-dimensional oscillator.

Let us suppose now that for an arbitrary operator $\hat{A}$ we have its Weyl symbol
\begin{equation}
f_A(\bm{q},\bm{p}) = \tr \left(\hat{A} \hat{U}(\bm{q},\bm{p}) \right),
\end{equation}
where the dequantizer operator $\hat{U}(\bm{q},\bm{p})$ is given by
\begin{equation}
\hat{U}(\bm{q},\bm{p})= (2\pi)^n  \hat{D}(\bm{q},\bm{p}).
\end{equation}
Then one has a reconstruction formula
\begin{equation}
\hat{A} = \int_{\mathbb{R}^{2n}} f_A(\bm{q},\bm{p}) \hat{D}(\bm{q},\bm{p}) \d \bm{q} \d \bm{p}.
\end{equation}
Moreover, the product of two operators $\hat{A} \hat{B}$ is mapped onto a star product of their symbols $f_A \star f_B$ whose explicit expression is
\begin{eqnarray}
\!\!\!\!\!\!\!\!\!\!\!\!\!\!\!\!\!\!(f_A \star f_B)(\bm{q},\bm{p})\!\! &=&\!\! \int_{\mathbb{R}^{4n}}\!\! f_A(\bm{q}_1,\bm{p}_1) f_B (\bm{q}_2,\bm{p}_2) 
\nonumber\\
&G &\!\!\!(\bm{q}_1,\bm{p}_1,\bm{q}_2,\bm{p}_2,\bm{q},\bm{p}) \d \bm{q}_1 \d \bm{p}_1 \d \bm{q}_2 \d \bm{p}_2,
\end{eqnarray}
where the Groenewald kernel is given by~\cite{MarmoJPA}
\begin{eqnarray}
& & G(\bm{q}_1,\bm{p}_1,\bm{q}_2,\bm{p}_2,\bm{q}_3,\bm{p}_3) 
\nonumber\\
& &= \tr \left(\hat{D}(\bm{q}_1,\bm{p}_1) \hat{D}(\bm{q}_2,\bm{p}_2) \hat{U}(\bm{q}_3,\bm{p}_3)\right)
\nonumber\\
& & = \frac{1}{\pi^{2n}} \exp\Big( 2\ii (\bm{q}_1 \cdot \bm{p}_2 - \bm{q}_2\cdot \bm{p}_1)  +  \text{cyc.\ perms.} \Big) \quad
\label{eq:rel12}
\end{eqnarray}
Let us now introduce for any operator $\hat{A}$ its symbol as a function on the manifold $X$
\begin{equation}
\mathcal{W}_A (x) = \tr\left(\hat{A} \hat{\varphi}(\hat{\bm{q}}, \hat{\bm{p}}, x) \right).
\label{eq:rel14}
\end{equation}
In view of relations~(\ref{eq:rel1})-(\ref{eq:rel2}) one can reconstruct the operator $\hat{A}$ using the formula
\begin{equation}
\hat{A}= \int_X \mathcal{W}_A (x) \hat{\chi}(\hat{\bm{q}}, \hat{\bm{p}}, x) \d x.
\label{eq:rel15}
\end{equation}
The relations~(\ref{eq:rel14}) and (\ref{eq:rel15}) mean that the operators $\hat{\varphi}(\hat{\bm{q}}, \hat{\bm{p}}, x)$ and $\hat{\chi}(\hat{\bm{q}}, \hat{\bm{p}}, x)$ play the role of dequantizer and quantizer, respectively. Thus, they determine the star-product of the symbols of operators $\hat{A}$ and $\hat{B}$ by
\begin{equation}
\mathcal{W}_A \star \mathcal{W}_B (x)\! = \!\int_{X^2} \mathcal{W}_A(x_1) \mathcal{W}_B (x_2) K(x_1,x_2, x) \d x_1 \d x_2.
\end{equation}
Therefore, starting from the classical integral relations~(\ref{eq:rel1})-(\ref{eq:rel2}) we constructed a new star-product scheme. In this scheme
the dequantizer and the quantizer are given by~(\ref{eq:rel3}) and (\ref{eq:rel4}), respectively. By the general theory of star-product description~\cite{MarmoJPA} the kernel of this star-product reads
\begin{equation}
K(x_1,x_2,x_3)\!=\! \tr(\hat{\chi}(\hat{\bm{q}}, \hat{\bm{p}}, x_1) \hat{\chi}(\hat{\bm{q}}, \hat{\bm{p}}, x_2) \hat{\phi}(\hat{\bm{q}}, \hat{\bm{p}}, x_3) .
\label{eq:rel19}
\end{equation}
By plugging the explicit expression~(\ref{eq:rel3}) and (\ref{eq:rel4}) into~(\ref{eq:rel19}) and using the definition of the Groenewald kernel~(\ref{eq:rel12}), we get
\begin{eqnarray}
K(x_1,x_2,x_3)&=& \int_{\mathbb{R}^{6n}} \chi(\bm{q}_1,\bm{p}_1, x_1) \chi(\bm{q}_2,\bm{p}_2, x_2) 
\nonumber\\
& & \varphi(\bm{q}_3,\bm{p}_3, x_3)
G(\bm{q}_1,\bm{p}_1,\bm{q}_2,\bm{p}_2,\bm{q}_3,\bm{p}_3)  
\nonumber\\
& & \d\bm{q}_1 \d\bm{q}_2 \d\bm{p}_2 \d\bm{q}_3 \d\bm{p}_3 .
\label{eq:rel20}
\end{eqnarray}
One application of formula~(\ref{eq:rel20}) is the tomographic scheme of star-product. For example, in the case of symplectic tomography the classical (Radon) transform and its inverse provide the functions (for simplicity we consider the single-mode case)
\begin{equation}
\phi(q,p,x) \equiv \phi(q,p, X,\mu,\nu) = \delta(X - \mu q - \nu p),
\end{equation}
with $x=(X, \mu, \nu)$, and 
\begin{equation}
\chi(q,p,x) \equiv \chi(q,p, X,\mu,\nu) = \frac{1}{4 \pi^2} e^{(\ii (X - \mu q - \nu p))}.
\end{equation}
Then the dequantizer and the quantizer of the tomographic star-product scheme read
\begin{equation}
\hat{\varphi}(\hat{q},\hat{p},x) = \delta(X- \mu \hat{q}-\nu \hat{p}),
\end{equation}
and
\begin{equation}
\hat{\chi}(\hat{q},\hat{p},x )  = \frac{1}{4 \pi^2}\exp \Big(\ii (X - \mu \hat{q} - \nu \hat{p})\Big),
\end{equation}
respectively. Applying the relation~(\ref{eq:rel20}) with the Groenewald kernel we find the kernel of the tomographic star-product obtained in~\cite{MarmoJPA}.

Now we apply the developed method to find the star-product kernel of a tomographic scheme based on quadratic functions~\cite{tomocurved}. Again we consider the single-mode example. We have
\begin{equation}
\varphi(q,p,x) 
= \delta\left(X-(q-\mu)^2-(p-\nu)^2\right),
\end{equation}
and
\begin{equation}
\chi(q,p,x) 
= \frac{1}{\pi}\exp\left(\ii(X-(q-\mu)^2-(p-\nu)^2)\right),
\end{equation}
with $x=(X,\mu,\nu)$.
The kernel of the quantum tomographic star-product  is given by the integral~(\ref{eq:rel20}), which in this case reads
\begin{eqnarray}
& & K(X_1,\mu_1,\nu_1,X_2,\mu_2,\nu_2,X_3,\mu_3,\nu_3) \nonumber\\
& & = \frac{1}{\pi^6}\int_{\mathbb{R}^{6}} \d{q}_1 \d{q}_2 \d{p}_2 \d{q}_3 \d{p}_3
\nonumber\\
& & \delta\left(X_3 - (q_3-\mu_3)^2 - (p_3-\nu_3)^2\right)
\nonumber\\
& &  \exp \left(\ii (X_1-(q_1-\mu_1)^2 - (p_1-\nu_1^2))\right) 
\label{eq:rel27} \\
& &  \exp \left(\ii (X_2-(q_2-\mu_2)^2 - (p_2-\nu_2^2))\right) 
\nonumber\\
& &  \exp \left(2\ii (q_1 p_2-q_2 p_1+q_2p_3-q_3 p_2 + q_3 p_1 - q_1 p_3)\right) .
\nonumber\\ \nonumber
\end{eqnarray}
A lengthy but straightforward calculation yields
\begin{eqnarray}
& & K(X_1,\mu_1,\nu_1,X_2,\mu_2,\nu_2,X_3,\mu_3,\nu_3)= \nonumber\\
& &\quad =\frac{2}{\ii \pi^3} \e^{\ii(X_1+X_2)} \e^{-\ii \frac{(\mu_1-\mu_2)^2+(\nu_1-\nu_2)^2}{2}}
\nonumber\\
& &\quad \delta\Big(4 X_3 - (\mu_1+\mu_2- 2\mu_3+\nu_2-\nu_1)^2 
\label{eq:rel28}\\
& &\quad \qquad\qquad -(\nu_1+\nu_2- 2\nu_3+\mu_1-\mu_2)^2 \Big)\nonumber.
\end{eqnarray}
The result is not symmetric with respect to the permutation $1\leftrightarrow 2$. This reflects the noncommutativity of the operator product. The result can be easily generalized to the multimode case.

\medskip

\section{Conclusions}

To conclude we point out the main results of our work. We have presented a general method to obtain the star-product of symbols of quantum observables in cases where the classical functions on phase-space are mapped onto tomographic symbols by means of different integral transforms. The method generalizes the Radon transform of functions on classical phase-space. The Radon transform is based on the use of a Dirac delta-function of a linear form in position and momentum. We used Weyl quantization as a tool to map functions on classical phase-space onto operators acting on a Hilbert space. By using this map we constructed the explicit expression of the quantizer and the dequantizer operators, providing the quantum version of tomography based on the Dirac delta-function of quadratic forms of position and momentum.   This form might be useful in the analysis of experiments for photon number measurement statistics.
The explicit expression of the star-product kernel of the generalized tomographic symbols was obtained in the case of quantum tomography determined by quadratic Hamiltonians. The generic relation of different star-product kernels with the  Groenewald kernel of the symbols of Weyl-Wigner functions on phase space was established.
The application of the developed tomographic star-product formalism will be the object of study of future work. It will be also interesting to  consider cases 
with discrete spectra like the angular momentum in a cylinder or systems with finite dimensional Hilbert spaces  like qudits where the star-product formalism can be 
easily implemented.

%%%%%%%%%%%%%%%%%%%%%%%%%%%%%%%%%%%%


\begin{thebibliography}{99}

\bibitem{NuovoCim}
M. A. Man'ko, V. I. Man'ko, G. Marmo, A. Simoni, F. Ventriglia,  Il Nuovo Cimento C 36 (2013) 163.

\bibitem{Ibort2}
A. Ibort, V. I. Man'ko, G. Marmo, A. Simoni, C. Stornaiolo, F. Ventriglia
Phys. Scr. {\textbf 88} (2013) 055003.


\bibitem{Rad1917}
J. Radon, \textit{\"Uber die bestimmung von funktionen durch ihre
integralwerte l\"angs dewisse mannigfaltigkeiten}, Breichte
Sachsische Akademie der Wissenschaften, Leipzig,
Mathematische-Physikalische Klasse, {\bf 69} S. 262 (1917).

\bibitem{Wig32}
E. P. Wigner, Phys. Rev. {\bf 40}, 749 (1932).

\bibitem{Ber-Ber}
J. Bertrand and P. Bertrand, Found. Phys. {\bf 17}, 397 (1987).

\bibitem{Vog-Ris}
K. Vogel and H. Risken, Phys. Rev. A {\bf 40}, 2847 (1989).

\bibitem{SBRF93}
D. T. Smithey, M. Beck, M. G. Raymer, and A. Faridani, Phys. Rev.
Lett. \textbf{70}, 1244 (1993).

\bibitem{lwry} A. I. Lvovsky and M. G. Raymer, Rev. Mod. Phys. \textbf{81}, 299 (2009).

\bibitem{Mlynek96}
J. Mlynek, Phys. Rev. Lett. \textbf{77}, 2933 (1996).

\bibitem{konst}
C. Kurtsiefer, T. Pfau, and J. Mlynek, Nature \textbf{386}, 150
(1997).

\bibitem{torino}
G. Zambra, A. Andreoni, M. Bondani, M. Gramegna, M. Genovese, G.
Brida, A. Rossi, and M. G. A. Paris, Phys. Rev. Lett. \textbf{95},
063602 (2005).

\bibitem{ps}  V. D'Auria, S. Fornaro, A. Porzio, S. Solimeno, S. Olivares and
 M. G. A. Paris,  Phys. Rev. Lett \textbf{102}, 020502 (2009).

\bibitem{bellini}  A. Zavatta, V. Parigi, M. S. Kim, H. Jeong and  M. Bellini,
Phys. Rev. Lett. \textbf{103}, 140406 (2009).

\bibitem{allevi09}
A. Allevi, A. Andreoni, M. Bondani, G. Brida, M. Genovese, M. Gramegna, S.
Olivares, M. G. A. Paris, P.  Traina and G. Zambra, Phys. Rev. A \textbf{80},
 022114 (2009).

\bibitem{MarmoJPA}
O. V. Man'ko, V. I. Man'ko, G. Marmo, J. Phys. A: Math. Gen. \textbf{35}, 699 (2002).

\bibitem{tomothick}
M. Asorey, P. Facchi, G. Florio, V. I. Man'ko, G. Marmo, S. Pascazio, E. C. G. Sudarshan,
Phys. Lett. A \textbf{375}, 861 (2011).

{\bibitem{quantomogram}
M. Asorey, P. Facchi, V.I. ManÕko, G. Marmo, S. Pascazio, E.C.G. Sudarshan,
Physica Scripta \textbf{85},  065001(2012).}

\bibitem{tomogram}
M. Asorey, P. Facchi, V.I. Man'ko, G. Marmo, S. Pascazio and
E. C. G. Sudarshan, Phys. Rev. A \textbf{76}, 012117 (2007).

\bibitem{tomocurved}
M. Asorey, P. Facchi, V. I. Man'ko, G. Marmo, S. Pascazio and
E. C. G. Sudarshan,
Phys. Rev. A \textbf{77}, 042115 (2008).


\end{thebibliography}
\end{document}